# Max-Planck-Institut
# Für Extraterrestrische Physik

astro-ph/9503012   2 Mar 1995

# ROSAT HRI Observations of the Crab Pulsar
# An Improved Temperature Upper Limit
# for PSR 0531+21

W. Becker and B. Aschenbach

Max-Planck-Institut für extraterrestrische Physik

85740 Garching bei München, Germany

# ROSAT HRI OBSERVATIONS OF THE CRAB PULSAR:
# AN IMPROVED TEMPERATURE UPPER LIMIT FOR PSR 0531+21

W. BECKER AND B. ASCHENBACH

*Max-Planck-Institut für extraterrestrische Physik,*
*85740 Garching bei München*
*InterNet: web@mpe-garching.mpg.de*

**Abstract.** ROSAT HRI observations have been used to determine an upper limit of the Crab pulsar surface temperature from the off-pulse count rate. For a neutron star mass of 1.4 $M_\odot$ and a radius of 10 km as well as the standard distance and interstellar column density, the redshifted temperature upper limit is $T_r^\infty \leq 1.55 \times 10^6$ K ($3\sigma$). This is the lowest temperature upper limit obtained for the Crab pulsar so far. Slightly different values for $T_i^\infty$ are computed for the various neutron star models available in the literature, reflecting the difference in the equation of state.

## 1. Introduction

Associated with the historical supernova of AD 1054, the Crab pulsar PSR 0531+21 is the youngest neutron star of a population of over 550 known rotation powered pulsars, and the only one, for which the age is known with high accuracy. The question about the thermal properties of the Crab pulsar has therefore been of special importance for the pulsar and neutron star theories since its discovery; knowing the surface temperature or temperature upper limit allows a comparison with the predictions of cooling theories for a neutron star at an early stage of its evolution and may yield important constraints for the equations of state and the internal structure of the neutron star. However, although the Crab is one of the most frequently observed sources in high energy astrophysics, previous experiments have failed to detect any spectral evidence for thermal emission from the neutron star's surface. The intense magnetospheric emission during the pulse-on phase prevents the detection of the thermal radiation associated with the neutron star's initial heat content, and during the pulse-off phase the thermal emission from the neutron star surface is buried under the bright synchrotron emission of the surrounding nebula. Only upper limits of the Crab pulsar surface temperature $T_s$ could be determined for that reason. First results in this respect were reported by Wolff et al (1975) and Toor & Seward (1977). Based on rocket born lunar occultation observations of the Crab supernova remnant in 1974, $T_s \leq 4.7 \times 10^6$ K and $T_s \leq 3 \times 10^6$ K, respectively, were inferred from the off-pulse count rate upper limit, which, interpreted in terms of blackbody radiation from the entire neutron star, is equivalent to a limit of the neutron star surface temperature. An improved result became available from the Einstein Observatory using the High Resolution Imager (HRI) few years later. Assuming a column density of $3 \times 10^{21}$ cm$^{-2}$ and a distance of 2 kpc, a $3\sigma$ upper limit of $T_s \leq 2.5 \times 10^6$ K was reported by Harnden & Seward (1984) for a neutron star of radius R=10 km and mass M=1.4 $M_\odot$. The lowest temperature upper limit re-ported for the Crab pulsar so far, was derived from an analysis of the 1969 and 1975 glitches and post-glitch behaviour and the interpretation in terms of unpinning and repinning of crustal superfluid vortex lines. The creep relaxation time obtained from a fit to the post-glitch timing data implies an internal temperature of $\sim 3 \times 10^8$ K, which for a 1.4 M$_\odot$ neutron star with a reasonable stiff equation of state results in a redshifted surface temperature upper limit of $T_s^\infty \leq 1.6 \times 10^6$ K (Alpar et al 1985).

In the present work, ROSAT HRI data from the Crab SNR are used to revisit the surface temperature of PSR 0531+21. Since the performance of the ROSAT X-ray telescope and HRI exceeds that of the Einstein Observatory in terms of the HEW angular resolution and scattering (Aschenbach 1988, David et al 1992), the pulsar can be separated more clearly from the surrounding nebula emission. The result is an improved upper limit, which for the first time constraints the thermal evolution models of neutron stars.

## 2. Observations and Data Analysis

The Crab SNR was one of the first targets observed in the ROSAT pointing program between 20-24 March 1991. During 8,785 s effective exposure with the HRI in the focus of the XRT, $\sim 2.35 \times 10^6$ cts were collected within a 2 arcmin wide circle centred on the pulsar. About 95% of the total emission was found to come from the intense synchrotron nebula (cf. Rosat Calendar 12/1992). To determine the off-pulse count rate from the DC level of the Crab pulsar's light curve, 111,366 counts were selected from a 6 arcsec wide aperture around the pulsar. After barycentring and correcting the photon arrival times using the approach described in Becker et al (1993a), each X-ray photon arrival time was related to the pulsar phase $\phi$, using the Crab Pulsar Timing ephemeris of 15. March 1992 (Lyne & Pritchard 1992). The corresponding Crab pulsar light curve in the ROSAT energy band is shown below.

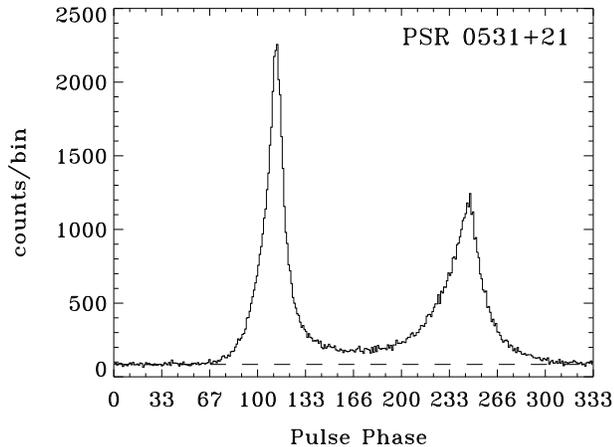

**Fig.1** Folded X-ray light curve of the Crab pulsar from ROSAT HRI observations of 20-24 March 1991. A single cycle is plotted in 333 phase bins of $\sim 100\,\mu$sec width. The DC level is fitted to $(84.5 \pm 1.2)$ cts/bin, indicated by the dashed line.

The (0.1-2.4) keV DC flux, integrated over the observation time, was found to be $(84.5 \pm 1.2)$ cts/bin by fitting a constant level to the off-pulse part of the X-ray light curve, shown in figure 1. However, the ROSAT HRI point spread function contains only 72% of the total energy from a point source within 6 arcsec radius by which the DC flux has been corrected to $(84.5 + 1.2) \times 100/72 \leq 119$ cts/bin. Since the X-ray image during the off-pulse phase does not show any evidence for a residual point-like source, the counts observed during this period are likely to be associated with the nebula rather than with the pulsar. Therefore, only an upper limit of 0.069 cts/s ($3\sigma$) for the Crab pulsar off-pulse flux can be derived (Becker 1994).

## 3. Results

Assuming blackbody emission from the surface of the neutron star of radius $R$, as well as the distance $d$ and absorption column density $N_H$, the surface temperature $T_s$ can be derived from the observed count rate. However, due to the gravitational redshift $z$ only $T_s^\infty = T_s/(1 + z)$ can be observed. For $R$=10 km, $M$=1.4 $M_\odot$, $N_H = 3 \times 10^{21}$ cm$^{-2}$ and $d$=2 kpc the $3\sigma$ upper limit of 0.069 cts/s corresponds to a $3\sigma$ temperature upper limit of $T_s^\infty = 1.55 \times 10^6$ K or $T_s = 2 \times 10^6$ K. This result is in line with the upper limit derived by Alpar et al (1985) and is about $0.5 \times 10^6$ K lower than the limit reported by Harnden & Seward (1984). The general relation between the redshifted surface temperature and the neutron star radius or bolometric luminosity $L_{bol}^\infty$, respectively, is shown in figure 2 for different interstellar column densities $N_H$.

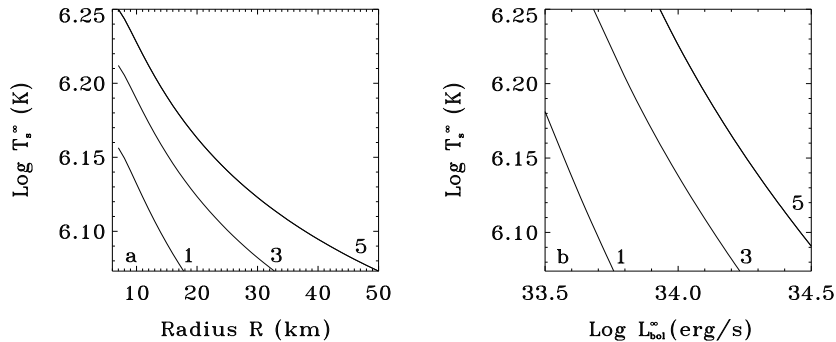

**Fig. 2a** Neutron star surface temperature $T_s^\infty$ vs. $R = R_\infty/(1 + z)$, inferred from the Crab pulsar off-phase count rate upper limit for $N_H = (1, 3, 5) \times 10^{21}$ atoms/cm$^2$. (**b**) Bolometric luminosity vs. $T_s^\infty = T_s/(1 + z)$. A distance of $d = 2$ kpc and $M$=1.4 $M_\odot$ have been assumed.

In the literature various neutron star models of different equations of state have been proposed. To compare our result with the predictions from the corresponding cooling curves, the stellar parameters $M$ and $R$ have been used for computing $T_s$ and $T_s^\infty$, respectively. Table 1 shows the results derived for the different models which are discussed, for example, in Nomoto & Tsuruta (1987), Van Riper (1991), Umeda et al 1993, Chong & Cheng (1993) and Page (1994). The errors quoted for $T$ and $L$ correspond to the uncertainty of $d$=($2 \pm 0.5$) kpc and $N_H = (3 \pm 1) \times 10^{21}$ atoms/cm$^2$.

TABLE 1   $3\sigma$ Temperature Upper Limits for PSR 0531+21

| Model | Radius | Mass | $T_s \times 10^6$ | $T_s^\infty \times 10^6$ | $\log T_s^\infty$ | $\log L_\gamma^\infty$ |
|-------|--------|------|------------------|--------------------------|-------------------|------------------------|
|       | km     | $M_\odot$ | K             | K                        | K                 | erg/s                  |
| PS[1]        | 16.10 | 1.31 | $1.61^{+0.19}_{-0.21}$ | $1.40^{+0.18}_{-0.16}$ | $6.15^{+0.05}_{-0.06}$ | $33.97^{+0.19}_{-0.24}$ |
| PS[2]        | 15.83 | 1.40 | $1.63^{+0.19}_{-0.21}$ | $1.40^{+0.18}_{-0.16}$ | $6.15^{+0.05}_{-0.06}$ | $33.97^{+0.19}_{-0.24}$ |
| MPA[3,4]     | 12.45 | 1.40 | $1.81^{+0.21}_{-0.24}$ | $1.48^{+0.19}_{-0.17}$ | $6.17^{+0.05}_{-0.06}$ | $33.90^{+0.19}_{-0.24}$ |
| PAL33[4]     | 11.91 | 1.40 | $1.85^{+0.22}_{-0.24}$ | $1.49^{+0.19}_{-0.18}$ | $6.17^{+0.05}_{-0.06}$ | $33.89^{+0.19}_{-0.24}$ |
| UV14[4,5]    | 11.20 | 1.40 | $1.90^{+0.23}_{-0.25}$ | $1.51^{+0.20}_{-0.18}$ | $6.18^{+0.05}_{-0.06}$ | $33.87^{+0.19}_{-0.24}$ |
| UU[6]        | 11.14 | 1.40 | $1.91^{+0.23}_{-0.25}$ | $1.51^{+0.20}_{-0.18}$ | $6.18^{+0.05}_{-0.06}$ | $33.87^{+0.19}_{-0.24}$ |
| PAL32[4]     | 11.02 | 1.40 | $1.92^{+0.23}_{-0.25}$ | $1.52^{+0.20}_{-0.18}$ | $6.18^{+0.05}_{-0.06}$ | $33.87^{+0.19}_{-0.24}$ |
| FP[1]        | 10.90 | 1.29 | $1.90^{+0.23}_{-0.25}$ | $1.53^{+0.20}_{-0.18}$ | $6.18^{+0.05}_{-0.06}$ | $33.85^{+0.20}_{-0.24}$ |
| FP[4,5]      | 10.85 | 1.40 | $1.93^{+0.23}_{-0.25}$ | $1.52^{+0.20}_{-0.18}$ | $6.18^{+0.05}_{-0.06}$ | $33.86^{+0.19}_{-0.24}$ |
| AV14[4,5]    | 10.60 | 1.40 | $1.96^{+0.23}_{-0.26}$ | $1.53^{+0.20}_{-0.18}$ | $6.18^{+0.05}_{-0.06}$ | $33.86^{+0.20}_{-0.24}$ |
| AU[6]        | 10.40 | 1.40 | $1.98^{+0.24}_{-0.26}$ | $1.53^{+0.20}_{-0.18}$ | $6.19^{+0.05}_{-0.06}$ | $33.85^{+0.20}_{-0.24}$ |
| FP(pion)[1]  | 9.40  | 1.27 | $2.04^{+0.25}_{-0.27}$ | $1.58^{+0.21}_{-0.19}$ | $6.20^{+0.05}_{-0.06}$ | $33.82^{+0.20}_{-0.24}$ |
| PAL(kaon)[7] | 8.20  | 1.40 | $2.27^{+0.27}_{-0.30}$ | $1.60^{+0.21}_{-0.19}$ | $6.21^{+0.05}_{-0.06}$ | $33.80^{+0.20}_{-0.24}$ |
| BPS[1]       | 7.90  | 1.35 | $2.30^{+0.28}_{-0.30}$ | $1.62^{+0.21}_{-0.20}$ | $6.21^{+0.05}_{-0.06}$ | $33.79^{+0.20}_{-0.25}$ |
| BPS[2]       | 7.35  | 1.40 | $2.45^{+0.30}_{-0.32}$ | $1.62^{+0.21}_{-0.20}$ | $6.21^{+0.05}_{-0.06}$ | $33.78^{+0.20}_{-0.25}$ |
| PAL(kaon)[7] | 7.00  | 1.40 | $2.55^{+0.31}_{-0.34}$ | $1.63^{+0.21}_{-0.20}$ | $6.21^{+0.05}_{-0.06}$ | $33.78^{+0.20}_{-0.25}$ |

The stellar parameters Radius and gravitational Mass are taken from: [1] Umeda et al (1993), [2] Van Riper (1991), [3] Müther et al (1987), [4] Page (1994), [5] Wiringa et al (1988), [6] Chong & Cheng (1993), [7] Thorsson et al (1994).

The comparison between the predicted cooling curves of the PS model and the surface temperature upper limit derived for the Crab pulsar is shown in figure 3.

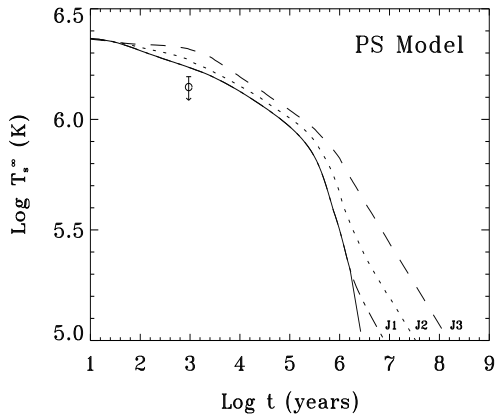

**Fig.3** $3\sigma$ surface temperature upper limit for PSR 0531+21, using the PS model parameters. Also shown are the cooling curves from Umeda et al (1993). Dotted and dashed lines represent the thermal evolution with internal frictional heating for strong (J3), weak (J2) and superweak (J1) pinning of crustal superfluid vortex lines.

## 4. Conclusion

Due to the superior performance of the ROSAT telescope an improved upper limit of the Crab pulsar surface temperature has been obtained. Whereas the previous upper limit derived from data of the Einstein Observatory was consistent with the prediction of standard neutron star cooling, even if heating processes like frictional heating or crust cracking have been taken into account (Van Riper 1991, Page 1992, Umeda et al 1993, Chong & Cheng 1993). Even more, c.f. figure 3, the temperature upper limit obtained with ROSAT is well below the temperature predicted by the PS model which invalidates its application. A comparison with the FP model shows agreement with the observation but only if the uncertainty in the distance and in the column density is stretched to the limits. Strong frictional heating in the FP model is also excluded by our observation.

A more detailed comparison between the different thermal evolution models and the Crab pulsar surface temperature upper limit is in preparation (Becker 1994).

## REFERENCES

Alpar, M.A., Nandkumar, R., Pines, D.: (1985), *Astrophys. J.* **288**, 191-195
Aschenbach, B.: (1988), *Applied Optics* No. 8 **27**, 1404-1413
Becker, W., Brazier, K., Trümper, J., (1993a), *Astron. Astrophys.* **273**, 421-424
Becker, W., Trümper, J. and Ögelman, H.B.: (1993b), in *Isolated Pulsars*, (eds
    K.A. Van Riper, R. Epstein & C. Ho), 104-109, (Cambridge University Press)
Becker, W.: (1994), PhD thesis, Ludwig-Maximillian-Universität München
Chong, N., Cheng, K.S.:(1993), *Astrophys. J.* **417**, 279-286
David, L.P. et al.: (1992), ROSAT HRI Technical Appendix
Harnden, F.R., Seward, F.D.: (1984), *Astrophys. J.* **283**, 279-285
Müther, H.,Prakash, M., Ainsworth, T.L.:(1987),*Phys. Lett.* **B119**, 469-473
Nomoto, K., Tsuruta, S.: (1987), *Astrophys. J.* **312**, 711-726
Page, D.: (1994), preprint, to appear in *Astrophys. J.* at June 10 1994
Prakash, M., Ainsworth,T.L., Lattimer, J.M.:(1988),*Phys. Rev. Lett.* **61**, 2518
Thorsson, Prakash, M., Lattimer, J.M.: (1994), preprint
Toor, A., Seward, F.: (1977), *Astrophys. J.* **216**, 560-564
Umeda, H., et al.: (1993), *Astrophys. J.* **408**, 186-193
Van Riper, K.: (1991), *Astrophys. J. Supp.* **75**, 449-462
Wiringa, R.B., Fiks, V., Fabrocini, A.:(1988), *Phys. Rev.* **C38**, 1010-1037
Wolff, R.S., et al.: (1975), *Astrophys. J.Lett.* **202**, L77-L81